# Optimization of Software Quality using Management and Technical Review Techniques

**Inibehe Emmanuel Akpannah**

Post Graduate Student (MSc. Information Technology), SRM University, Chennai, India

***Abstract***

Optimizing the quality of software is a function of the degree of reviews made during the early life of a software development process. Reviews detect errors and potential errors early in the software development process. The errors detected during the early life cycle of software are least expensive to correct. Efficient involvement in software inspections and technical reviews, help developers improve their own skills, thereby mitigating the occurrence of errors in the later stage of software development process. The ideas gathered on this paper point that a properly implemented program of technical and management reviews drastically reduces the time as well as the cost required for testing, debugging, and reworking, and dramatically improves the quality of the resulting product. This paper, Optimization of Software Quality using management and technical Review Techniques, provides its readers with the opportunity to learn about and experience using this indispensable software quality tools.

## Introduction

In software, either as a process or a product, quality is essentially known to be an important factor. In other to achieve greater improvement in the quality and productivity of a software process, it is important to perform design and code reviews. In software development process, a small coding error can result in a critically vulnerability that ends up compromising the security of the entire system. Improving the quality of our software process leads to minimal rework, cost effectiveness and also meeting up with schedules, which all lead to improvement in capability measurement. Software review is a chain process, it is dynamic and review needs to be performed at every phase of the development process until there is a transition to the testing phase.

Software design and code review is a phase in the software development process in which programmers (authors of codes), quality assurance (QA) team, and peer reviewers get together to review design and code [9]. One way of optimizing the quality of software is finding and correcting errors at the early stage of a software development process. Finding and fixing errors at the early stage of a software development process is relatively inexpensive and tends to minimize the more expensive process of managing, identifying and fixing defects during later stages of development or even after the software are delivered to the users.

Reviewers who are responsible for reading software codes read the code line by line to check for:

- present flaws and potential flaws
- The correctness and uniformity of the overall program design.
- Validity of comments, some comments may result into misinformation
- Adherence to the generally accepted coding standards.

A software product, during the process of its development, developers should never assume it to be free of flaws. Employing review techniques will limit the number of coding errors and thereby reducing the degree of impact it will have on the software.

## Problem Definition

The underlying problem is the question of how to use existing review techniques (management and technical reviews) to optimize the quality of software. Although there are other techniques of optimizing the quality of software during its development process other than review, the goal of this paper is to present how review techniques can be applied in a software development process in other to improve and optimize its quality on an economical basis.

## Contribution

This paper proposes a dynamic approach to evaluate the quality of software using review techniques based on the opinion of experts and predicted behaviour of the software. It shows a more cost effectiveness over testing at the later stage of the software development process. Based on reliability models, it is seen that the quality of a software product is higher than a similar product that has not been reviewed.

## Types of Review

The Institute of Electrical and Electronic Engineers (IEEE) Standard for software Reviews defines five types of Review [3]:

- Management Review
- Technical Review
- Inspections(Formal Peer Review)
- Walk-through
- Audits

## Management Reviews

According to IEEE, in Software Engineering, Management Review is defined as a systematic evaluation of a software acquisition, supply, development, operation, or maintenance process performed by or on behalf of management to monitor progress, determine the status of plans and schedules, confirm requirements and their system allocation, or evaluate the effectiveness of management approaches used to achieve fitness for purpose [8]. Management reviews support decisions about corrective actions, changes in the allocation of resources, or changes to the scope of the project.
Management review is performed by those directly responsible for the system. They monitor the progress of the system and determine status of plans and schedules. The management team confirms requirements and their system allocation or, evaluates management approaches used to achieve fitness or purpose. Management review requests that support decisions are made about changes to the scope of the project. Software developers may at a given point determine to change or modify the scope of the project, the support for this decision comes directly from the management. The management reviews the progress of the program and evaluates why changes and/ or modifications should be made. If there is a strong reason for alteration, the management produces a statement of support otherwise it is not allowed. It is different from both a software engineering peer review which evaluates the technical quality of software products, and a software audit, which is an externally conducted audit into a project's compliance to specifications, contractual agreements, and other criteria.

## The Aim of Management Review

The aim of Management Review is to manage the quality of software and of its development process. A quality product is one which meets its requirements and satisfies the user requirements. A quality culture is an organizational environment where quality is viewed as everyone's responsibility.

## Products Reviewed by Management

The following are the products reviewed by management:

- plans Audit Reports
- Contingency
- Installation plans
- Risk management Plans
- Software quality Assurance(Q/A)
- Progress Report
- Software Project Management Plans
- Software Safety Plans
- Backup and Recovery Plans
- Technical Review Reports
- Software Product Analyses
- Verification and Validation Reports

## Outputs of Management Review

From the series of reviews carried out by the management, each activity reviewed is properly documented [1]. The document is helpful to the management in identifying: the project under review, review team members to check for their skills and technical know- how in case there is need for replacement of a team member, review objects and, review input and also list of defects identified by the review team.

## Technical Review

A software technical review is a form of peer review in which a team of qualified personnel examines the suitability of the software product for its intended use and identifies discrepancies from specifications and standards. Technical reviews may also provide recommendations of alternatives and examinations of various alternatives (IEEE Standards 1028 -1997, IEEE Standard for Software Review, Clause 3.7)[9].A technical review is a software quality assurance activity performed by software engineers with well-defined objectives.

## Objectives of Technical Review

Technical review offers the following objectives:

- To uncover errors in function, logic or implementation of the software ;
- To verify that the software meets its requirements ;
- To ensure that the software has been developed according to the standards ;
- To achieve uniformity in software development process;
- To make projects manageable.

The formal technical review serves to promote backup and continuity because a number of people become familiar with parts of the software that they may not have been seen. Each technical review is conducted as a meeting and is considered successful only if it is properly planned, controlled and attended.

### Software Products Reviewed Technically

Examples of software products reviewed technically include but not limited to the following:

- Installation Procedure
- Software Design Description
- Software Test Documentation

- Software User Documentation
- Release Notes
- System Build Procedure
- Maintenance Manual

**Factors that enhance Technical Review of Software**

There are some factors that spur the need for a technical review in software [9]:

- **Technological factors**: Underlying technologies may have some changes. For example the software that was initially thought to be useful or provide alternative measures may not come up as earlier envisaged. So doing a technical review will help to analyse the changes and thereby avoiding developing software of poor quality.
- **Man Power/Staffing factors**: Poor staffing can result into software with poor quality. To avoid this, it is recommended that staff with reputable record of software development experience and good understanding of programming languages are employed. Staffing plays a major factor in software development. 70% of the quality of a product directly or indirectly lies in the hands of the staff involved. Proper staff selection is advised.
- **Organizational factors**: There may be changes within the organization which is handling the software project. These changes can affect the quality of software.
- **Changing requirements**: There may be some changes in the requirements for the project. The client may decide to alter the requirements and this should be reviewed technically before implementation. Everyone involved in the process should be aware of the changes in other to prevent unnecessary costs of testing and amendments.
- **Compliance with standards and best practices**: It may be necessary to ensure that the project has implemented quality assurance processes to ensure that project deliverables comply with appropriate standards and best practices.

## Approaches to a Technical Review

Discussed below are the basic approaches to a technical review:

- **Team Review**: The project development team may wish to reflect on the approaches they have taken. It is expected of the development team to make reports to the project manager stating the progress of the project.
- **Review by project partners**: The project partners are also granted the responsibility of reviewing the project.
- **Invitation of third parties:** The project team may wish to invite external bodies to participate in the review. This approach is very dynamic. It reviews the project in various dimensions.
- **Comparison with one's peers**: One may also chose to compare one's deliverables with one's peers, such as projects with close similarity.

## Outputs of a Technical Review

It is important to note that any improvements or changes which may have been identified during a review need not necessarily be implemented. There may be a temptation to implement best practices when good practices are sufficient, and that implementation of best practices may take longer to implement than anticipated. The outputs from a review may be:

- **Better understanding**: The review may have an educational role and allow project partners to gain a better understanding of issues.
- **Enhanced workflow practices**: Rather than implementing technical changes the review may identify the need for improvements to workflow practices.
- **Documenting lessons**: The review may provide an opportunity to document limitations of the existing approach. The documentation could be produced for use by project partners, or could be made more widely available.
- **Deployed in other areas**: The recommendations may be implemented in other areas which the project partners are involved in.
- **Implemented within project**: The recommendations may be implemented within the project itself. If this is the case it is important that the change is driven by project needs and not purely on technical grounds. The project manager should normally approve significant changes and other stakeholders may need to be informed

## Compile and Test Time Range

Figure 1.0 shows the maximum, minimum and average percentage of total development time that six engineers in one project spent compiling and testing the software. Initially the engineers averaged around 40 percent of their time in compile and test, with the percentage of their development time ranging from around 10 percent to over 50 percent. As the engineers worked through the project the quality of their programs improved substantially. By introducing reviews and being more aware of the defects they made, the engineers cut their average compile and test time by nearly four times [3].

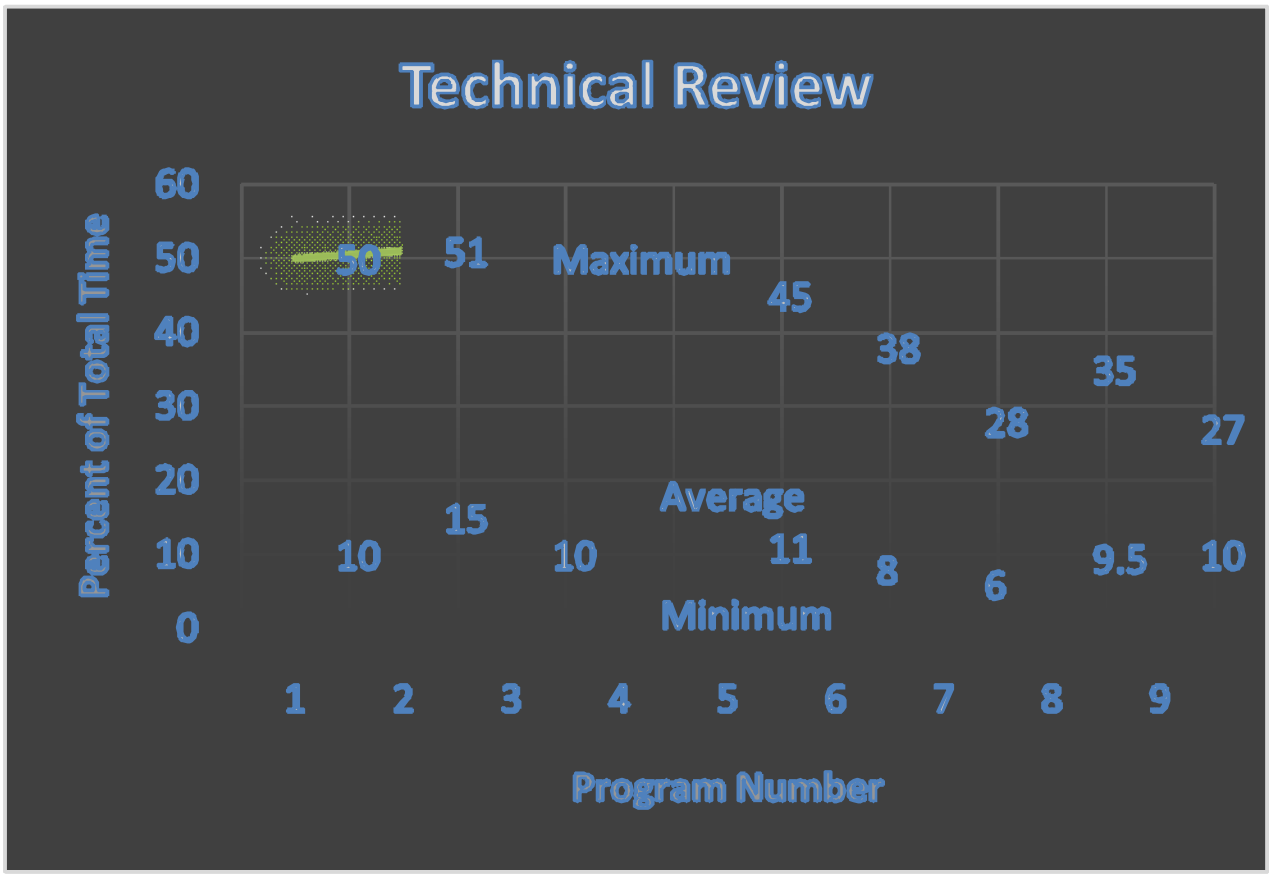

**Figure 1.0 Range for Compile and Test Time.**

## Conclusion

The purpose of software review is to ensure that programs produced are of the highest quality and thereby optimized to achieve that state of quality. Of the many kinds of reviews available today, the major ones are the management and technical reviews, they are the backbones and gangways to other reviews. The optimization of software quality starts by reviewing the requirements, design, documentation and perhaps any other product element. Doing the review of all these elements contributes greatly to ascertaining the quality of our software product.

## Future Enhancement

Practitioners can make of these review methods and in combination of other methods that have not been explicitly explained here to further improve on the quality of software. The technical review can be extended to industrial scale and more analyses can predict a more optimized way of improving the quality of software products other than management and technical reviews.

## Acknowledgement

The author hereby acknowledges the grace of God upon his life for his devotion towards this work. Also special thanks go to his favorite lecturers for their encouragement and inspiration: S. Kanchanna, K.Kanmani.